\documentclass[aip,jcp,preprint,amsmath,amssymb,endfloats*]{revtex4}
\usepackage{graphicx}% Include figure files
\usepackage{epstopdf}
\usepackage{amsmath}
\usepackage{dcolumn}% Align table columns on decimal point
\usepackage{bm}% bold math

\makeatletter 
\def\@dotsep{4.5} 
\makeatother

\newcommand{\BE}{\begin{equation}}
\newcommand{\EE}{\end{equation}}
\newcommand{\f}{\frac}
\newcommand{\bdm}{\begin{displaymath}}
\newcommand{\edm}{\end{displaymath}}
\def\r{\rangle}
\def\l{\langle}
\begin{document}

%\preprint{To be submitted to  J. Chem. Phys. }

\title{Equilibrium properties of a  grafted polyelectrolyte with explicit counterions} 
\author{Kandiledath Jayasree}
\email{jayasree@physics.iitm.ac.in}
\affiliation {Department of Physics, Indian Institute of Technology Madras, Chennai - 600 036, India}

\author{P. Ranjith}
\email{pranjith@curie.fr}
\affiliation {Physico-Chimie UMR 168, Institut Curie, 26 rue d'Ulm,  75248 Paris Cedex 05, France}

\author{Madan Rao}
\email{madan@ncbs.res.in}
\affiliation{Raman Research Institute,  Bangalore - 560 080, India\\
and National Center for Biological Sciences (TIFR), GKVK Campus, Bangalore - 560 065, India}
  
\author{P. B. Sunil Kumar}
\email{sunil@physics.iitm.ac.in}
\affiliation {Department of Physics, Indian Institute of Technology Madras, Chennai - 600 036, India}

\date{\today}

\begin{abstract}
{We study the equilibrium conformations 
of a grafted polyelectrolyte (PE) in the presence of explicit counterions (CI) using Monte Carlo simulations. The interplay between attractive Lennard-Jones  interactions (parametrized by 
$\epsilon$), and electrostatics (parametrized by $A=q^2 l_B/a$, where $q=$counterion valency, $l_B=$ Bjerrum length and $a=$ monomer diameter), results in a variety of conformations, characterized as extended ($E$), 
pearls with $m$ beads ($P_m$), sausage ($S$) and globular ($G$).  For large $\epsilon$, we observe a  transition from  $G  \rightarrow P_2 \rightarrow P_3 \rightarrow \ldots  \rightarrow S \rightarrow G $ with increasing $A$, {\it  i.e.}, a change from  poor to good, to {\it reentrant} poor solvent,
whereas, at lower $\epsilon$, the sequence of transitions is,  $E \rightarrow S \rightarrow G$. 
The conformation changes are directly related to the nature of binding of  CI onto the
PE. The transition between $S \to G$ is continuous and associated with critical fluctuations in the
shape driven by fluctuations in the fraction of condensed CI.}
\end{abstract}

%\pacs{87.16.-b, 64.75.+g, 68.05.Cf}

\maketitle

\section{INTRODUCTION}

 The equilibrium properties of a polyelectrolyte (PE) are governed by an interplay between long-range attractive forces between monomers, 
electrostatic interactions arising predominantly from mobile counterions (CI), and polymer elasticity. Of these forces, electrostatics has been the most difficult to handle, both analytically and computationally, especially 
in the regime where the strength of the electrostatic interaction,   
parametrized by the ratio of 
the electrostatic condensation energy of the multivalent counterions and 
thermal energy, is high. While the behaviour at high temperature or low electrostatic energy 
is well described by the mean field
Poisson-Boltzmann (PB) theory~\cite{brilliantov-98, lee-04,muthu-04}, several computer simulation studies on PE, starting from~\cite{deserno-00},
have reported qualitative deviations from PB at low temperatures or high electrostatic energy~\cite{deserno-00,stevens-93,winkler-98, dunweg99,holm-01,holm-02,muthu-02,netz,wada}.

These studies reveal that the main cause for the deviation, is that PB theory underestimates  the extent of CI condensation at high electrostatic coupling and low temperatures, and neglects spatial correlations. For instance, a study of the the equilibrium conformations of a PE as a function of charge density and solvent  quality (restricting to weak electrostatics and poor solvent conditions) using a Debye-Huckel framework~\cite{dunweg99}, demonstrated that with increasing charge density, the PE globule splits into a string of pearls, in agreement with scaling arguments~\cite{dobrynin96}. 
Simulations with explicit 
CI~\cite{muthu-02} verified the above and further showed a 
 collapse of the PE at higher electrostatic coupling, due to attractive, dipolar interactions arising from the condensation of CI onto the monomers.  Apart from these studies on free PE,  the conformations of a constrained PE, such as a grafted PE, in the regime of low electrostatic coupling were
 examined using molecular dynamics (MD) simulations~\cite{crozier-03}. 
 
 In this paper, we study the equilibrium phase diagram of a grafted PE, across a range of electrostatic couplings and solvent quality,  using Monte Carlo (MC) simulations  and scaling arguments. Our study  highlights the competition between the monomer-monomer,
monomer-counterion interactions and polymer elasticity. As a consequence, the PE exhibits a variety of phases, which we characterize as extended ($E$), 
pearls with $m$ beads ($P_m$), sausage ($S$) and globular ($G$).  With increasing  electrostatic interaction, the PE exhibits the following sequence of conformations :  $G  \rightarrow P_2 \rightarrow P_3 \rightarrow \ldots  \rightarrow S \rightarrow G $. Thus, as a function of increasing
electrostatic interaction, we go from poor to good to reentrant poor solvent. These conformation changes are intimately tied to the nature of binding of CI onto the PE. For weak electrostatic interactions, the CIs  condense onto the PE and partially screen the
monomer charge. However when the electrostatic interactions are stronger, we find that new composite degrees of freedom, such as dipoles comprising of monomer and  CI charges, emerge. In addition, we find that the  transition between $S$ to the reentrant  $G$ is continuous and associated with critical fluctuations in the
shape driven by fluctuations in the fraction of condensed CI.

 \section{Model and Simulation details}
Our model PE is a linear chain of $N$ spherical beads,
of charge ${\it e}$ and diameter ${\it a}$, connected through harmonic springs.
One end of the PE is anchored to the  wall  at $x=0$.
To ensure charge neutrality, we introduce $N$ oppositely  charged counterions with the same valency and diameter. We neglect hydrodynamic effects and treat the solvent  as a dielectric continuum with permittivity $\kappa$.  The system is bounded within a cubic box of volume $L^3$, with  impenetrable, non-polarizable walls at $x=0$ and 
$x=L>aN$. We apply periodic minimum image boundary conditions along the $y$ and $z$ directions~\cite{netz}. We ignore any contribution from image charges at the boundaries along the $x-$axis.

In our simulation, we model the non-electrostatic interactions by (i) a Lennard-Jones (LJ) potential,
\begin {equation}
U(r) = \left\{ \begin{array} {ll}
\epsilon_1\sum_{i<j}\f{a ^{12}}{({\bf r}_i-{\bf r}_j)^{12}}-2\f{a ^6}{({\bf r}_i-{\bf r}_j)^6}  & \,\,\,|\bf{r}_i-{\bf r}_j|<a,  \\
\\
\epsilon_2\sum_{i<j}\f{a ^{12}}{({\bf r}_i-{\bf r}_j)^{12}}-2\f{a ^6}{({\bf r}_i-{\bf r}_j)^6} &\,\ |\bf{r}_i-{\bf r}_j| \ge a,
\end{array}
\right .
\end{equation}
between all particles and (ii) a harmonic spring potential  acting between connected beads of the PE,
\BE
U_{s}=\sum_{i}k(|{\bf r_i}-{\bf r}_{i+1}|-a)^2,
\EE
where we take the spring constant, $k=800 k_BT/{a^2}$, for the simulation. For  monomer-monomer interaction we choose $\epsilon=\epsilon_1=\epsilon_2$
and for the CI-CI and monomer-CI interactions $\epsilon_1=1$ and $\epsilon_2=0$.
 The electrostatic potential between any pair of beads is,
\BE
U_c=A\sum_{i<j}\f{a s_is_j}{|{\bf r}_i-{\bf r}_j|} \, ,
\EE
where $s_i$ gives the sign of the ion charge. The electrostatic coupling strength  $A=\f{q^2 l_B}{a}$ measures the ratio of the
 coulomb to thermal energy when the distance of separation between two charges is  $a$; here $q$ is the valency of the monomer and $l_B=e^2/4\pi \kappa k_BT$, is the Bjerrum length.  From now on, we will write  all distances in units of  $a$ and energy in units of $k_BT$.

We determine the equilibrium conformations and phase diagram of this grafted, flexible PE, with explicit counterions, over a wide range of values  of the electrostatic coupling $A$ and  LJ parameter $\epsilon$, using the standard Monte Carlo Metropolis scheme. 
We use the well known Umbrella sampling technique~\cite{torrie74,frenkel92} to obtain the free energy profile of the 
PE-CI system as function of its radius of gyration, $R_g$. This method uses a weight function to bias
 the MC sampling of configuration space in such a way that the less probable states of the system are sampled frequently. 
In our simulation, we introduce the weight function by adding a harmonic potential $U_w=\f{1}{2}C{(R_g-R_0)}^2$ to the 
total energy of the system. By varying $R_0$, we sample the whole configuration space and measure the distribution of $R_g$, from which we calculate the free energy $F(R_g)$ of the PE.

\section{Equilibrium Phase Diagram}

\begin{figure}
\centering
\includegraphics[scale=0.38]{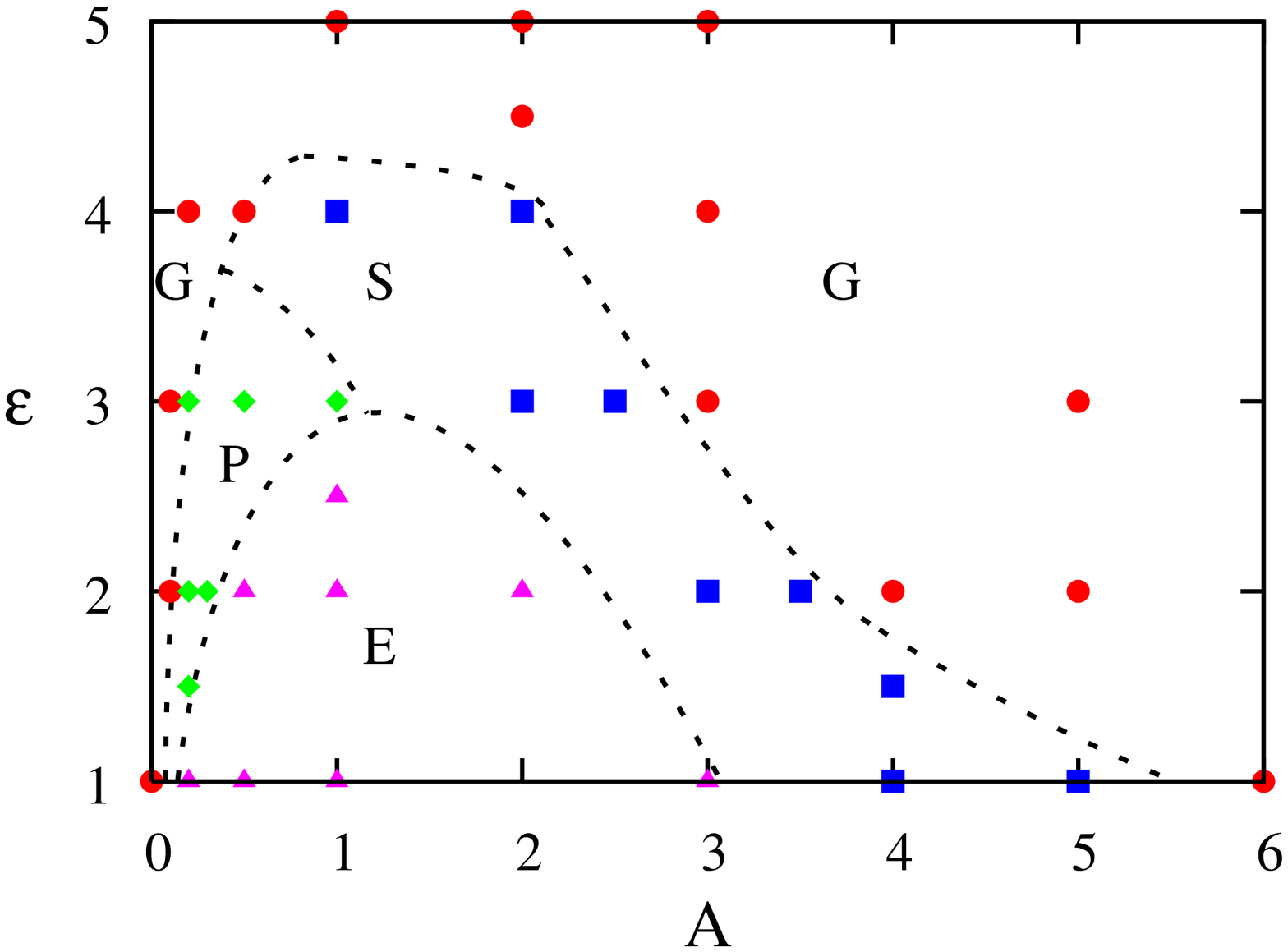}
\caption{Phase diagram of the PE in the $A-\epsilon$ plane for $N=120$. Different phases are ,(i) G: Globular, (ii) P: Pearls, (iii) S: Sausages (iv) E: Extended.}
\label{A-eps}
\end{figure}

\begin{figure*}[ht]
\begin{center}
\includegraphics[scale=0.4]{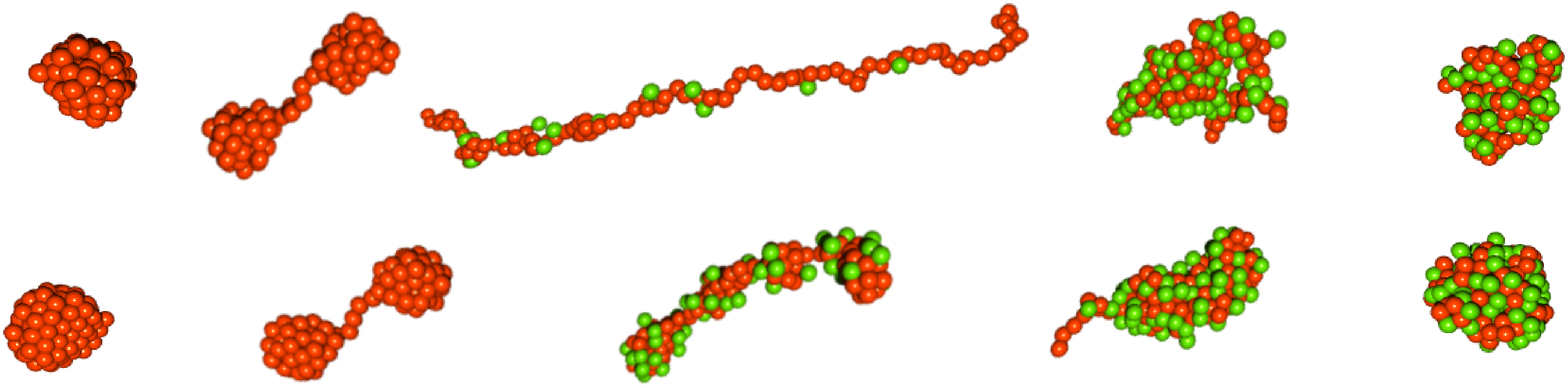}
\caption{ Snapshots of the  equilibrium configurations of the grafted PE (red) with condensed CI ( green) , typical of  the different phases.  {\em Top panel:} For  $N=100,\epsilon=2$, the configurations from the left are (i) $G$ : Globular at $ A=0.1$, (ii) $P_2$ : Pearls with 2 beads at  $A=0.2$, (iii) $E$: Extended at  $A=1$, (iv)$S$ : Sausage at $ A=3.5$, and (v) $G$ : Globular at $ A=4$. {\em Bottom panel:} For $N=120, \epsilon=3$, the configurations from the left are  (i) $G$ : Globular at  $A=0.1$, (ii) $P_2$ : Pearls with 2 beads at $A=0.2$, (iii) $P_4$ : Pearls with 4 beads at $ A=1$, (iv)$S$ : Sausage at $ A=2.5$ and (v) $G$ : Globular at $A=3$. }

\label{e3-n120}

\end{center}
\end{figure*}

The equilibrium phase diagram of the grafted 
PE, determined by an  interplay
between interactions (electrostatics and attractive LJ) and entropy (polymer and CI), is fairly subtle and is shown in Fig.~\ref{A-eps}.
We characterize the phases as Globular ($G$), Sausage ($S$), Pearls with $m$-beads ($P_m$)
and Extended (E), by the $N$-dependent scaling of conformational  measures, such as the
radius of gyration $\l R_g\r$ and the anisotropy of the gyration tensor;  typical conformations are shown in  Fig.~\ref{e3-n120}. Let us remind ourselves of  the polymer conformations when $A=0$. As a function of $\epsilon$, the polymer goes from being 
a self-avoiding random coil in good solvent conditions, characterized by the Flory behaviour, $\l R_g\r \sim N^{3/5}$, to a collapsed globule (G) in a poor solvent, characterized by $\l R_g\r \sim N^{1/3}$, via a first-order transition at $\epsilon=1$~\cite{degennes78}.

 \begin{figure}[htpb]
\centering
\includegraphics [scale=0.4]{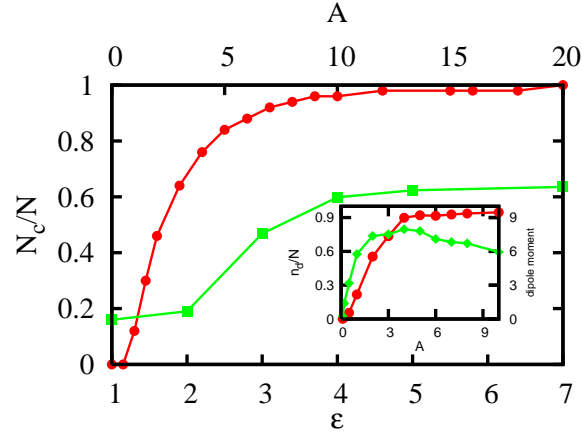}
\caption{ Fraction of condensed CI  as a function of (i) the LJ attraction $\epsilon$ (square) for $A=1$ and  (ii)  the electrostatic 
interaction $A$ (circles) with $\epsilon_1=1$ and $\epsilon_2=0$. A CI is defined as condensed if it is within a distance $2a$ from any  monomer of the PE. The inset shows the net dipole moment of the PE (square) and number of dipoles $n_d$ ( circles) as a function of $A$ for $\epsilon=2$.}
\label{eps-Nc}
\end{figure}

As we turn on the electrostatic interaction, the PE undergoes a sequence of shape changes, depending on the value of $\epsilon$. In the range, $\epsilon < 1$, the PE changes from a self-avoiding random walk (SAW)  to an extended conformation (E), as $A$ varies between $0$ and $1$ ~\cite{muthu-02,netz}. This is because, while the electrostatic interaction increases, the CI entropy prevents condensation onto the PE. The net 
electrostatic interaction between the monomers is repulsive resulting in an extension of the PE,
and an $\l R_g\r \sim N$.
When $A>1$, a finite fraction of  CIs condense on the monomers; at $A=5$, for instance, $80\%$  of CI condense onto the PE (Fig.~\ref{eps-Nc}). This results in 
reducing the net monopole charge on each monomer, and pairing the monomeric charge with the CI  to form dipoles ~\cite{muthu-02}. Such dipoles are the {\it emergent} degrees of freedom when $A \gg 1$; this leads to an effective attractive interaction, and the configuration resembles a Sausage (S). As $A$ is increased to $10$,  complete  condensation of the CI  takes place, resulting in a 
 collapse into a globule (G).

 For larger values of the attractive LJ potential, a new set of phases intervene between the 
globular phase (G) at $A=0$, and the extended phase (E). These phases occur when $\epsilon \gtrsim1$
(stronger short range attraction) and $A <1$ (negligible CI condensation), and therefore arise from a 
competition between short range attractive interactions, long range electrostatic repulsion between
monomers and polymer entropy, akin to the well-known Rayleigh instability. 
The resulting conformation is a string of pearls with $m$-beads ($P_m$), with $m$ increasing with $A$~\cite{kantor95,schiessel98,dobrynin96,dunweg99} (see Fig.~\ref{e3-n120} ). 
At large scales, the conformations in the $P_m$ phase is dominated by the electrostatic repulsion between the pearls,
and the radius of gyration scales as $R_g(N) \sim N$, with a prefactor proportional to the bead size.
The subsequent behaviour as a function of $A$ depends on the value of the attractive $\epsilon$. 
In the range, $1 < \epsilon  < 3$, the PE changes from $P_m$ (where $m \ll N$ is the maximal allowed by the finite
size of the polymer)
to an extended (E) conformation, as $A$ increases towards $1$.  This is because the attractive 
interactions are not strong enough to compete with the increasing electrostatic (monomer-monomer) repulsion as $A$ increases. In this range of $\epsilon$, the conformation then changes from
$E \to S \to G$, as described in Fig.~\ref{e3-n120}.  As we increase the attraction $\epsilon \gtrsim 3$, we lose the extended phase entirely, and the PE
goes from $G \to P_2 \to P_3 \to \ldots S$, and subsequently to a reentrant $G$ phase as we tune the 
electrostatic repulsion $A$ as shown in Fig.~\ref{e3-n120}.  

\begin{figure}[ht]
\begin{center}
\includegraphics [scale=0.35]{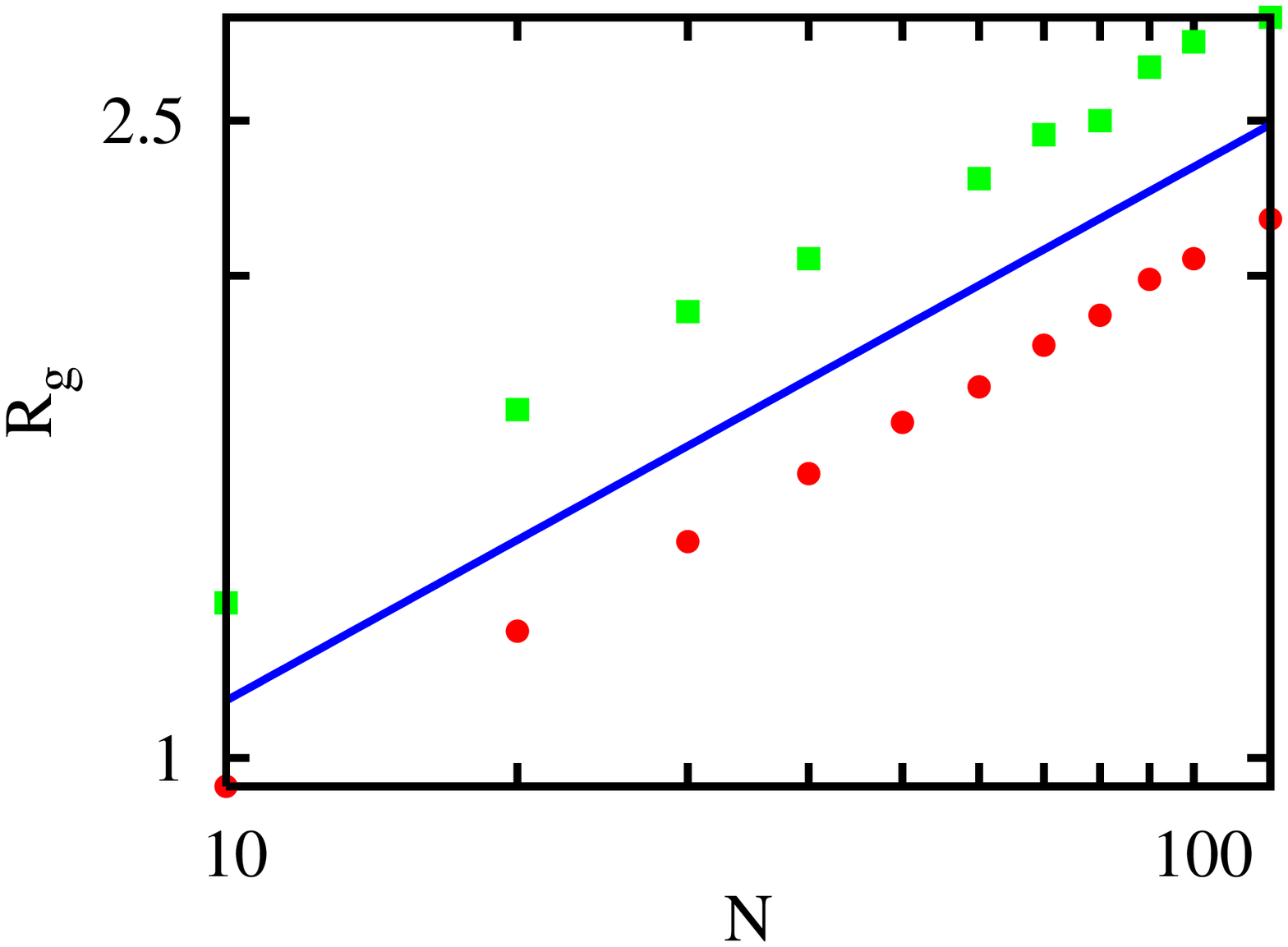}
\caption{  Scaling of the mean radius of gyration $\l R_g\r$ with $N$ for the globular configurations shown in Fig.~\ref{e3-n120} :  The  circles are for  $\epsilon=3.0, A=0.1$
and squares are for $\epsilon=3.0, A=3.0$.  The solid line is a guide to the eye with   $\l R_g\r \sim N^{1/3}$.   }
\label{scaling}
\end{center}
\end{figure}

 The conformations at large $A$, namely, the S and the reentrant G phases, are a consequence of the emergent dipole degrees of freedom.
 In the inset of Fig.~\ref{eps-Nc} we plot  the net dipole moment and the fractional number of emergent dipoles ($n_d/N$) of the PE as a function of $A$ for $\epsilon = 2$.  The dipole moment  shows  a maximum before the globular phase is reached -- at this stage, the number of dipoles is small. As we increase $A$, the number of dipoles increase, but the net dipole moment starts to decrease, as the PE conformation gets more compact in the G phase. The scaling of the radius of gyration in the two G-phases (at low and high $A$) is the same ---  $R_g(N) \sim N^{1/3}$ (poor solvent !), as shown in Fig.~\ref{scaling}.Thus, the reentrant G-phase, while structurally similar to the initial G-phase at $A=0$, is different only in its electric dipole characterisitics.

At larger values of the attractive potential $\epsilon \gtrsim 4$, the PE conformations go directly
from $G \to S \to G$ or even from $G \to G$ at  still higher values of $\epsilon \gtrsim 5$, where the latter 
$G$-phase has a dipole moment.  These direct transitions  are consistent with the behaviour of the fraction of condensed CI, which
{\it  increases} with  $\epsilon$, for a fixed $A$ (Fig.~\ref{eps-Nc}). The enhancement of CI condensation, which 
is a result of  electrostatic correlations,  results in a strong screening of the monomer-monomer interactions.

Before we end  this section, a word about finite size effects. In our simulations on finite PEs, the scale for the maximum bead number in the $P_m$ phase is set by system size, and in the thermodynamic limit of $N \to \infty$, there should be a finite fraction of beads. At this stage we are unsure about how the phase boundaries discussed above would shift as we increase $N$, or even whether the S phase exists in the thermodynamic  limit.

\section{Continuous transition from $S\to G$ : anomalous fluctuations of the sausage}

\begin{figure}
\centering
\includegraphics [scale=0.3]{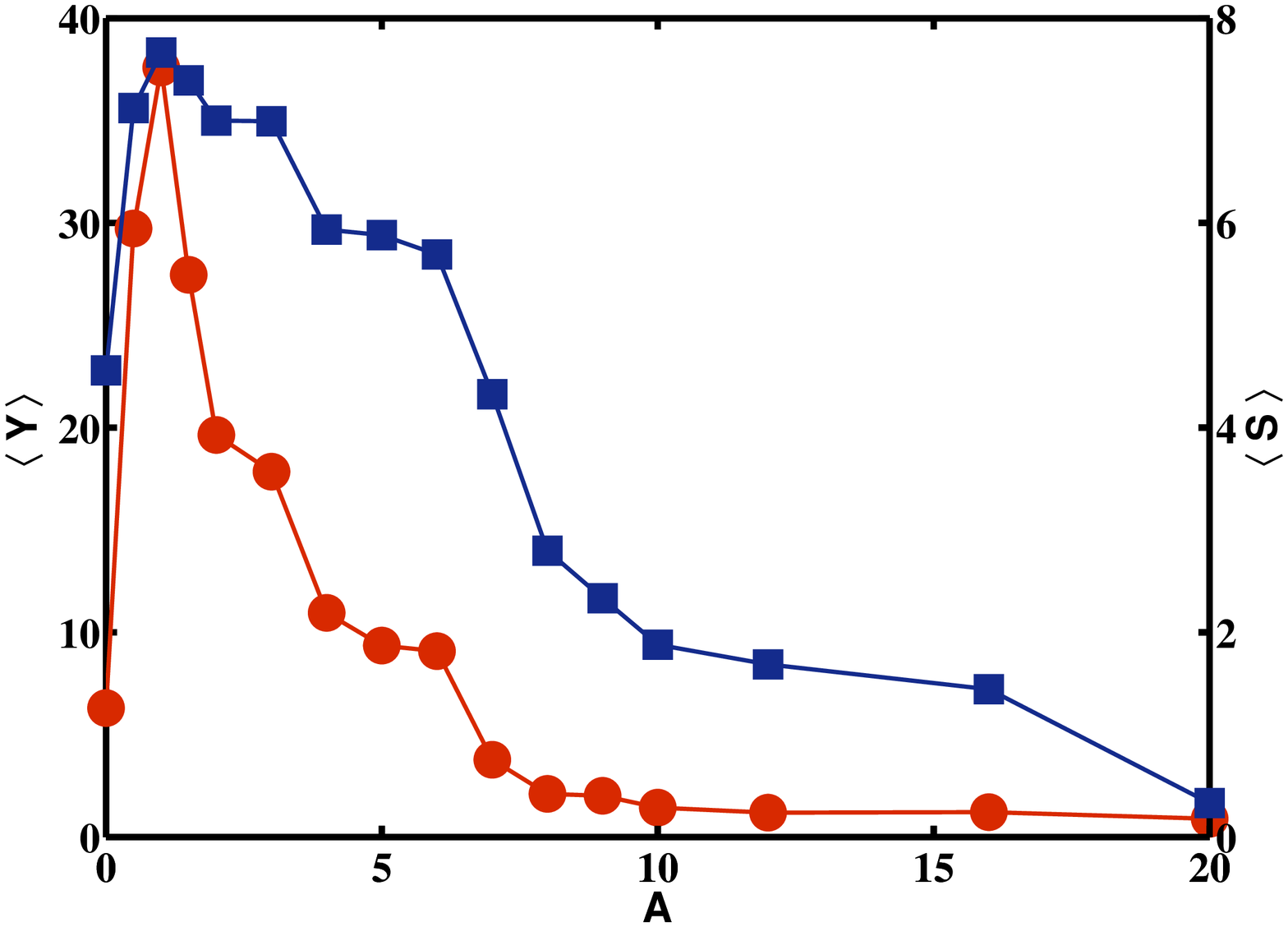}
\caption{Asphericity  $\l Y \r$ (circles) and $\l S \r =  \l R^2 \r/ \l R_g^2\r -2 $ (squares) (see text), as a function of electrostatic coupling $A, \epsilon_1=1.0,\epsilon_2=0.0$, for $N=50$. For reference, $Y=0$ for a globule, $\l S \r=4$ for a gaussian chain and $\l S \r=10$ for a rigid rod. }
\label{aspect}
\end{figure}

The phase diagram (Fig.~\ref{A-eps}) shows a transition from (reentrant) $G \to S$, upon {\it reducing} the electrostatic coupling $A$ 
from a high value. This transition is a symmetry breaking transition; the appropriate order parameter characterizing the spontaneous breaking of spherical symmetry is the {\it asphericity} parameter, $\l Y \r \equiv  2 \l \lambda_{1}^2 \r /( \l \lambda_2^2\r + \l \lambda_3^2 \r)-1$
where $\{\lambda_1, \lambda_2, \lambda_3\}$, are the eigenvalues of the  gyration tensor of the PE,
with $\lambda_1$ as the largest. The asphericity $\l Y\r=0$ for a globule;  Fig.~\ref{aspect} shows 
how this asphericity changes as a function of $A$. At the $G \to S$ transition, this symmetry breaking 
order parameter changes continuously, suggesting a 2nd-order phase transition. On the other hand
the $G\to E$ transition at small $A$ seems abrupt, suggesting a first-order transition.
We also plot the structural quantity $\l S \r \equiv  \l R^2 \r/ \l R_g^2\r-2 $, where $R$ is the squared end-to-end distance~\cite{csajka-00,netz03a}, as a function of $A$; this quantity is $\l S \r=4$ for a gaussian chain and $\l S \r=10$ for a rigid rod. 

\begin{figure}[htpb]
\centering
\includegraphics [scale=0.35]{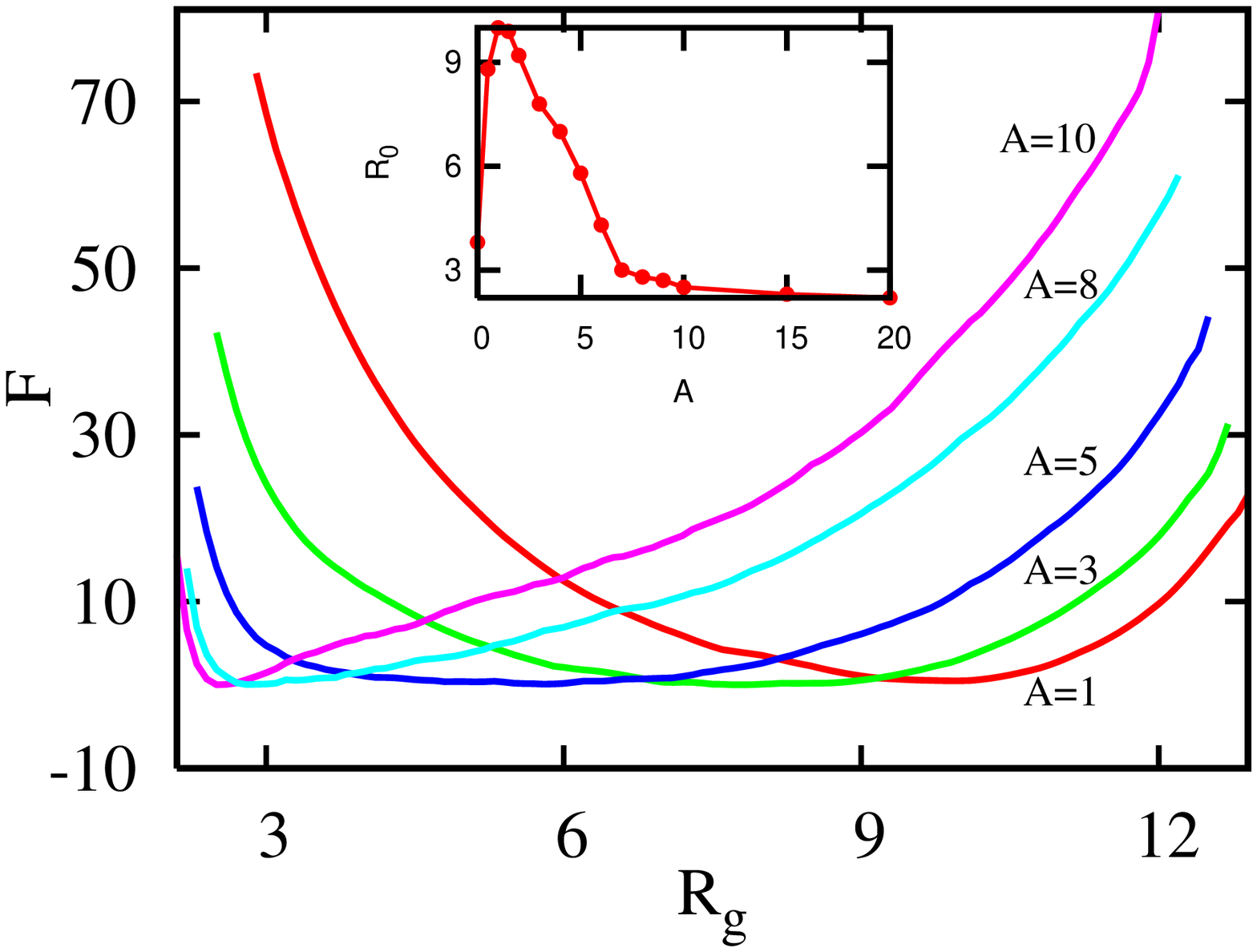}
\caption{ Free energy $F(R_g)$ for different coupling constants, of a  N=50 and L=50 PE, for $\epsilon_1=1.0$ and  $\epsilon_2=0.0$. (inset) $R_0$, the value of $R_g$ at which the free energy is  the minimum as a function of $A$, showing 
smooth change.}
\label{fenergy}
\end{figure}

To understand the nature of the phase transitions better, we compute the free energy of the PE with the CI, as a function of $R_g$,  for different values of $A$.   The smooth variation of the free energy profile and its minimum with increasing $A$ (Fig.~\ref{fenergy}), as one moves from $E\to S\to G$,
is consistent with the continuous transition described above. 
Indeed at around $A\sim 5$,  when we are in the $S$-phase, the minima in the free energy profile
 is extremely shallow, suggesting that the fluctuations of shape in this regime would be large,  consistent with its proximity to a critical point. 
 
\begin{figure}[htpb]
\centering
\includegraphics [scale=0.35]{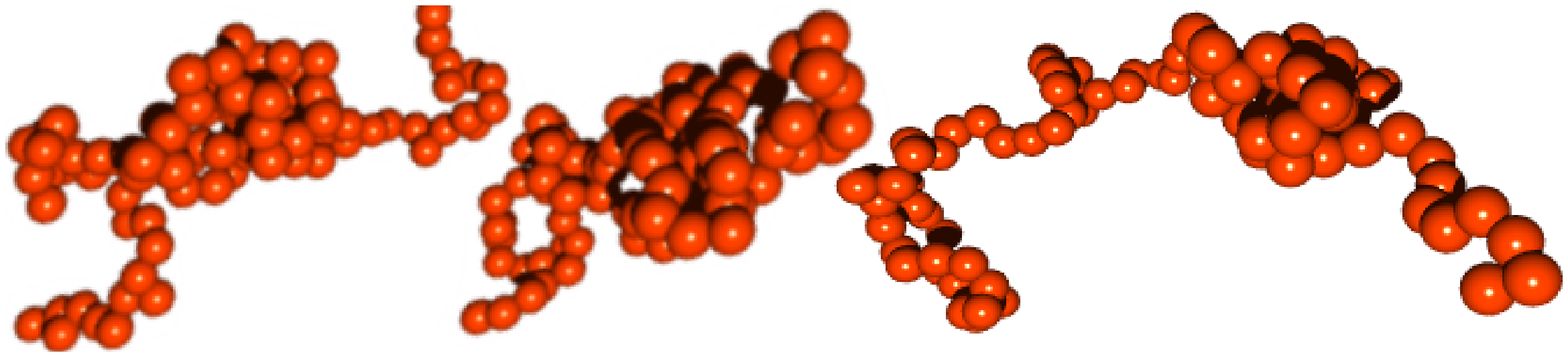}
\caption{ Configuration snapshots of the Sausage (S) showing strong fluctuations at equilibrium, for $A=5, \epsilon=1$. The configurations change from being extended to collapsed.}
\label{A-5-conf}
\end{figure}

\begin{figure}[htpb]
\centering
\includegraphics [scale=0.35]{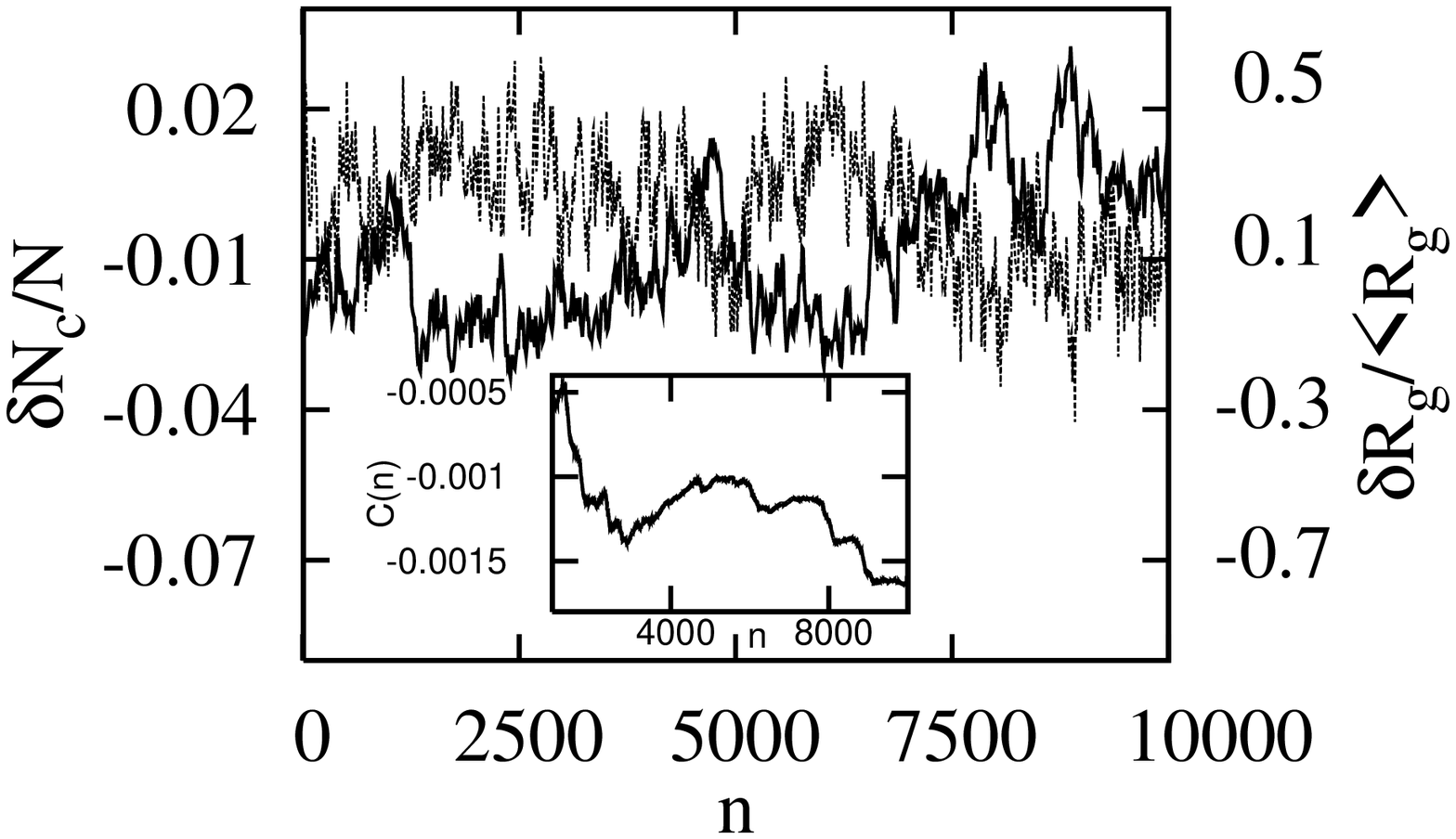}
\caption{ The  deviation of fraction of condensed CI from its average value $\delta N_c/N$ (dotted line),  and the deviation of $R_g/\langle R_g\rangle$ from its average value $\delta R_g/\l R_g \r$ (solid line) for different equilibrium configurations with $A=5$, $ \epsilon=1$ is shown.  The cross correlation $C(n) $ is shown in the inset} 
\label{ci-rg-cor}
\end{figure}

We explicitly study the fluctuations in the configurations in the S-phase; snapshots of the equilibrium
configurations displayed  in Fig.~\ref{A-5-conf}, show very strong shape fluctuations, with significant sampling of both extended and collapsed conformations.  
These large fluctuations of the PE shape in the S-phase are accompanied by
strong fluctuations in the fraction of condensed CI (Fig.~\ref{ci-rg-cor}).  
The dynamical interplay between  the PE shape and condensed CI ,
is shown in Fig.~\ref{ci-rg-cor} -- whenever there is a transient enhancement of  the condensed CI
fraction, the PE chain gets more compact (owing to the effective dipolar attraction), and whenever the condensed fraction is low, the chain gets stretched out (owing to strong monomer-monomer repulsion) leading to  a large negative value of the  cross-correlator $C(n)=\frac{1}{n}\sum_{i=1}^{n} (\delta N_c (i)\delta R_g(i))/( \l R_g \r N)  $, where $n$ is the number of  Monte Carlo steps . 

\begin{figure}[htpb]
\centering
\includegraphics [scale=0.3]{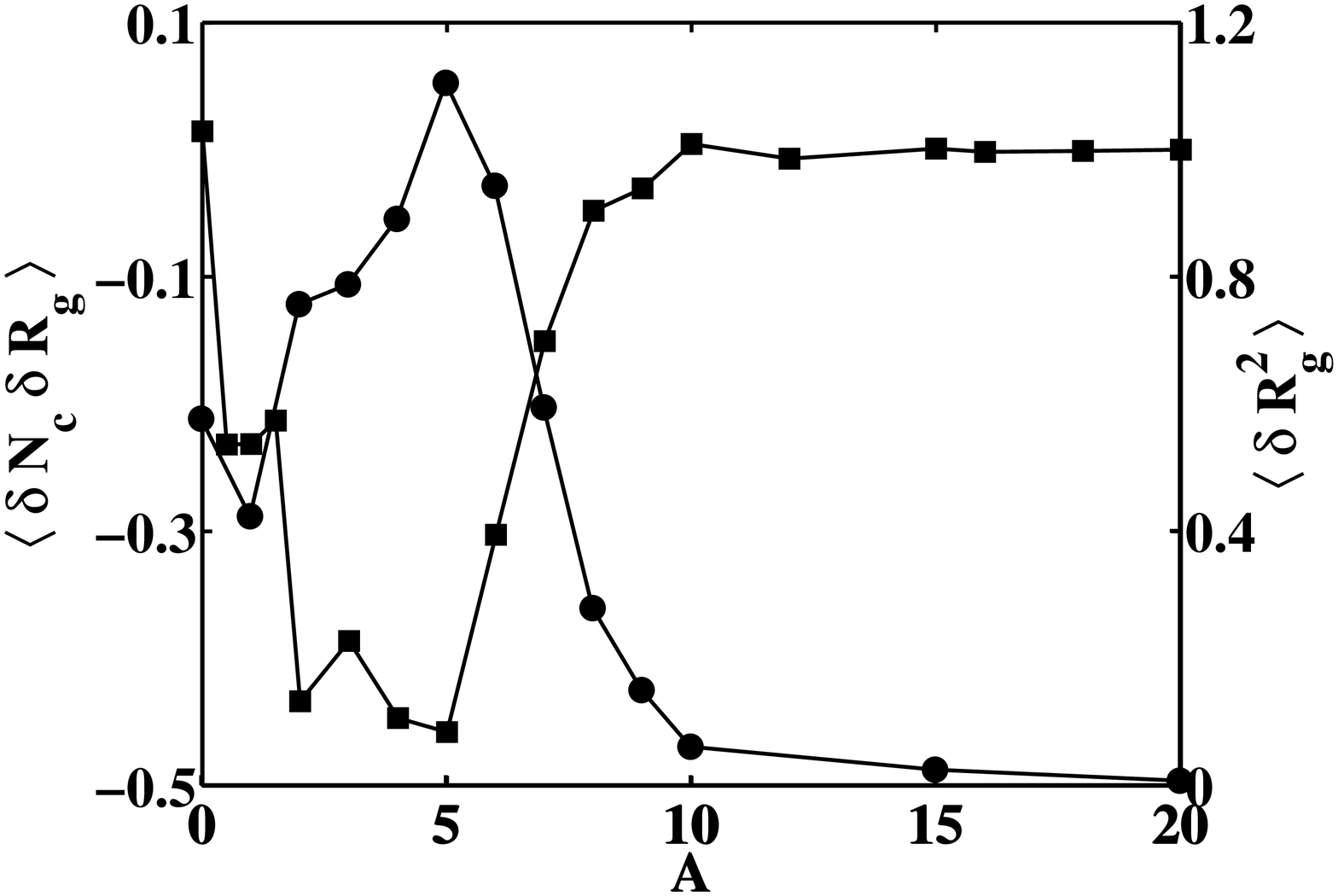}
\caption{ The   deviation of $R_g$ from its average value $\langle \delta R_g^2 \rangle$ (circle) and the cross-correlator $\langle \delta R_g \delta N_c\rangle$ (square) as a function of $A$ for $\epsilon_1=1.0,\epsilon_2=0.0$} 
\label{secondmoment}
\end{figure}

To quantify these fluctuations as a function of $A$   we compute the second moment of  $R_g$,  $\l \delta R_g^2\r = \l (R_g- \l R_g \r ) ^2\r $ , using the free energy $F(R_g)$. This quantity shown in Fig.~\ref{secondmoment}  exhibits  a peak around $A=5$ ( in the $S$ regime), with  its height increasing with system size, indicating  a critical point.
The  cross-correlator, $\l \delta R_g \delta N_c \r$ shows a negative peak at exactly where the peak in  $\l \delta R_g^2\r$ appears reiterating the interplay between PE shape and condensation of CI.  Our preliminary study to determine the order of the phase transition needs to be reinforced by a more detailed study of the scaling of fluctuations as a function of chain length $N$; we hope to return to this  when we have better computational facilities at our disposal. However, if we take our study as
 evidence of a continuous transition, then this would imply that upon increasing $\epsilon$, the critical line would necessarily terminate in a critical end point.

\section{Conclusion}

Using extensive Monte Carlo  simulations, we have explored the phase diagram of a grafted PE with explicit CIs, as a  function of the electrostatic coupling $A$  and LJ interaction parameter $\epsilon$. 
We have charaterized the phases in terms of the statistics of their conformations, and electrostatic measures, such as fraction of condensed counterions. We have uncovered  four distinct phases : globular, string of pearls with $m$-beads, extended, sausage and reentrant globule with nonzero dipole moment.

Our study, which treats the CI explicitly, highlights the strong correlation between collapse of the PE and the condensation of the CI. The dependence of CI condensation on $\epsilon$ (at low values of $A$),
the emergent formation of dipoles from condensation of CI onto the PE, the anomalous fluctuations of the sausage, are all a consequence of these correlated ion effects. The appearance of new degrees
of freedom, viz., dipoles, when $A>1$, makes it difficult to construct an analytical theory valid for all 
values of $A$. One possible approach would be  to represent the CI by  `two-species',  a
monopole charge density and a dipole density, whose relative fraction depends on $A$. We will explore these ideas in a future
submission.

\section*{Acknowledgments} 
The simulations were carried out  at the High Performance Computing Facility at IIT Madras. 
MR thanks HFSP and CEFIPRA 3504-2 for funding.

%\bibliographystyle{prsty}
%\bibliography{polyelec}

\listoffigures
\end{document}